\documentclass[doublecol]{epl2}
\usepackage{graphicx,amsfonts,subfigure,pifont,url}

\newcommand{\be}{\begin{equation}}
\newcommand{\ee}{\end{equation}}
\newcommand{\bea}{\begin{eqnarray}}
\newcommand{\eea}{\end{eqnarray}}

\title{Dynamics of gene expression and the regulatory inference problem}

\author{Johannes Berg}
\institute{
Institut f\"ur Theoretische Physik,
Universit\"at zu K\"oln \\ 
Z\"ulpicherstr. 77, 50937 K\"oln, Germany\\
and \\
Kavli Institute for Theoretical Physics,\\       
UCSB, Santa Barbara, CA 93106-4030
}


\abstract{
  From the response to external stimuli to cell division
  and death, the dynamics of living cells is based on the expression
  of specific genes at specific times.  The decision when to express a
  gene is implemented by the binding and unbinding of transcription
  factor molecules to regulatory DNA.  Here, we construct stochastic
  models of gene expression dynamics and test them on experimental
  time-series data of messenger-RNA concentrations.  The models are
  used to infer biophysical parameters of gene transcription,
  including the statistics of transcription factor-DNA binding and the
  target genes controlled by a given transcription factor. 
}

\begin{document}

\maketitle

\section{Introduction}

The primary step in the production of a protein is the transcription of
a gene from the DNA template to a m(essenger)-RNA molecule. Gene
transcription is carefully controlled, allowing the cell to respond to
situations requiring different proteins: for instance,
enzymes are produced when nutrients need to be processed and repair
proteins are assembled to respond to cell damage.

\begin{figure}[tbh!]
\centering
\subfigure[\label{fig:transcription_schematic}]{
\includegraphics[width=.36\textwidth]{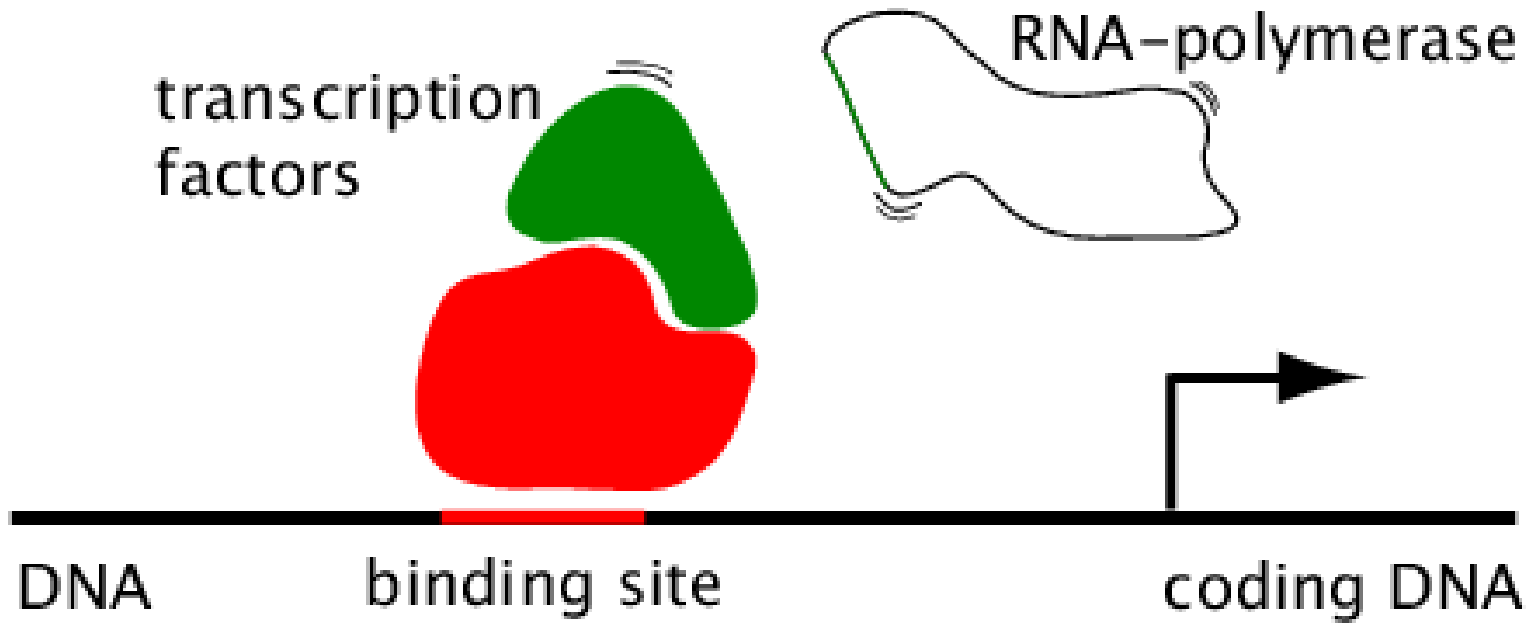}}
\subfigure[\label{fig:microarray_schematic}]{
\includegraphics[width=.42\textwidth]{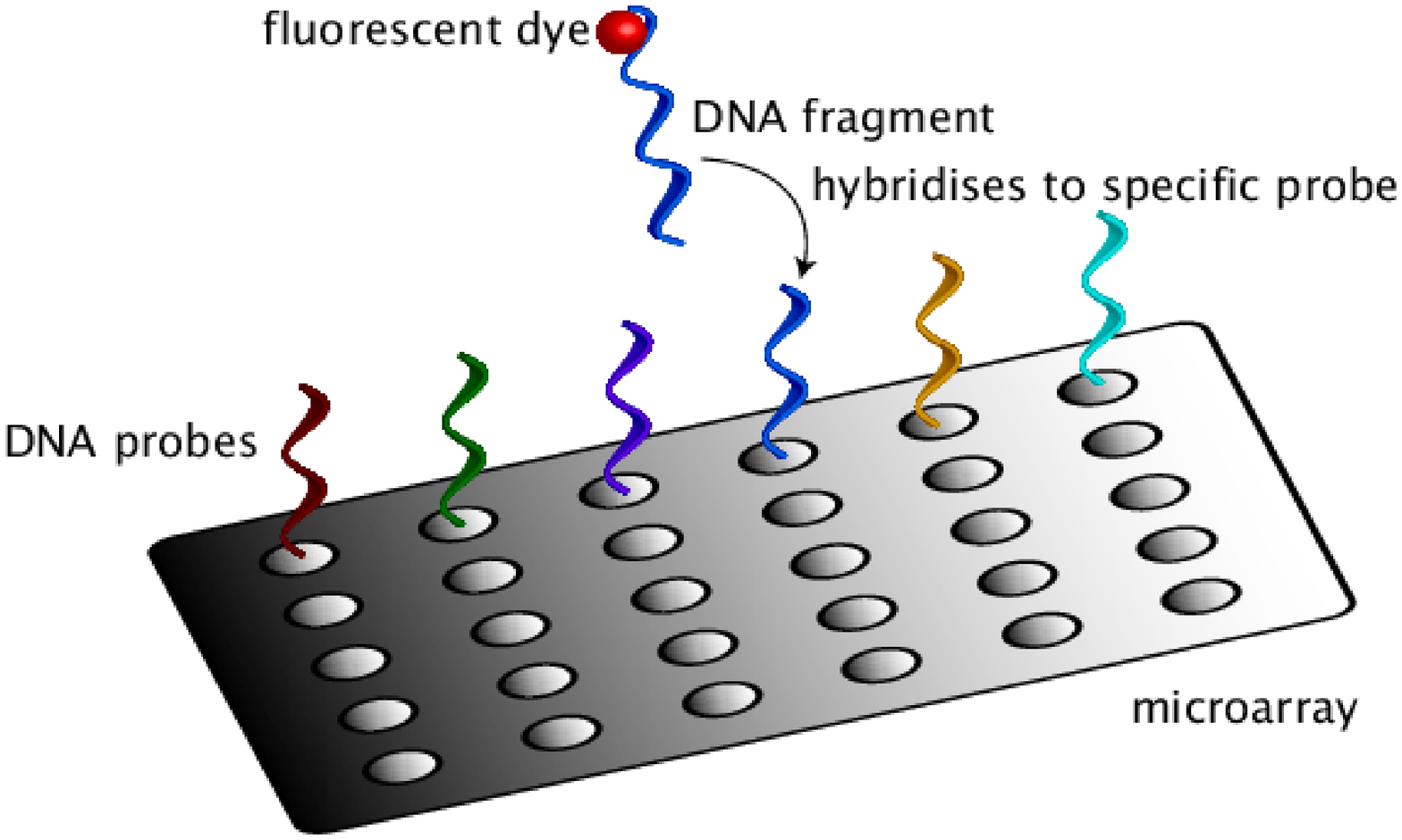}}
\caption{ \small {\bf Biophysics of gene transcription and its experimental 
measurement.} 
  a) Transcription factors are proteins which bind to specific sites 
on DNA. These binding sites are located in the so-called regulatory region
of a gene, which for yeast typically extends over several hundred
nucleotides before the starting point of transcription (marked with the arrow). 
A fully 
assembled set of transcription factors will attract and activate the machinery (called 
RNA-polymerase) which transcribes the DNA template to an mRNA molecule. In
a further step, the mRNA molecule will be translated to a protein. 
\newline
b) DNA-microarrays can simultaneously measure mRNA expression 
levels in a sample for all genes of an organism. A sample is taken
from a population of cells in a certain state, so even for low
expression levels of a gene, the number of molecules is macroscopic. 
(Single-cell fluctuations are averaged out in this process.)
mRNA in a sample is reversely
transcribed to DNA and deposited on a chip, which has a strand of
complementary DNA grafted onto its surface. The DNA in the sample will 
bind to its complement on the chip.
Fluorescent labelling of the DNA in the sample allows to determine the 
original mRNA concentration relative to a reference sample. By photolithographic techniques, hundreds 
of thousands of such DNA-probes can be placed on a single microarray-chip. 
By now over $10^5$ genome-wide experimental 
measurements of mRNA-expression levels have been performed for different 
organisms, under different experimental conditions, or for different 
tissues~\cite{GEOgene_expression}. }
\end{figure}

The cellular information processing to achieve this control has a 
concrete biophysical basis. Transcription 
factor molecules locate and bind to specific sites on DNA found in the
regulatory region in front of a gene. The transcription factors then
attract or repel the molecular machinery that reads off the gene (see
Fig.~\ref{fig:transcription_schematic}). The resulting gene product
may itself be a transcription factor, enhancing or suppressing the
transcription of further genes. The dynamics of gene expression is
thus governed by a complex web of molecular interactions. Changes in
the regulatory interactions (caused for instance by mutations in a 
transcription factor binding site) disrupt the cellular dynamics and 
can lead to serious defects, including cancer. 

Which transcription factor targets which gene can in principle 
be determined experimentally, by investigating the
physical binding of transcription factors to the regulatory DNA of a
target gene. High-throughput methods have resulted in lists of
potential target genes for many transcription
factors~\cite{Leeetal:2002,Harbisonetal:2004}, which, however, contain
many false positives and false negative results.
Binding sites for transcription factors can also be identified 
computationally by searching regulatory DNA for statistically 
overrepresented subsequences~\cite{HertzStormo:1999,BussemakerSiggia:2000}. 

Both approaches, however, tell nothing about the function
of a transcription factor (for instance its activity as an inhibitor
or an enhancer of gene transcription), or about its interplay with
other transcription factors. Moreover, some transcription factors
(called co-factors) do
not bind directly to DNA but to other transcription factors, and thus
do not have binding sites on DNA.  As a result, even in well-studied
organisms such as yeast, the function and target genes of about 
half the transcription factors remain unknown~\cite{Chuaetal:2004}.

The development of microarrays has made it possible to simultaneously
measure the expression levels (mRNA concentrations) of all genes of an
organism (see Fig.~\ref{fig:microarray_schematic}). This allows to
probe the response of the regulatory network to external
perturbations, or to detect correlations of expression levels of
different genes~\cite{Margolin.etal:2004}.

The effort to deduce the input-output relations of regulatory networks
from experimental gene expression data, however, is impeded by the
large number of such relations which are compatible with a given set
of expression levels~\cite{Hertz:1998}.  For instance, an observed
correlation between mRNA concentration of two genes may stem from a
causal relationship (one gene regulates the expression of the other),
or from both genes sharing a common
regulator~\cite{Margolin.etal:2004}. As a result, many machine
learning approaches to the network reconstruction problem, such as
Bayesian probabilistic models (see \cite{Friedman:2004,BarJoseph:2004} for reviews), or linear
regression~\cite{Lebre:2007} are well-known to be afflicted with the
curse of dimensionality~\cite{Hertz:1998,ChuGlymourScheinesSpirtes:2003}.
 
Here we take a biophysics approach to gene transcription. The
inference of the biophysical machinery underlying gene transcription
from its output is a challenging inverse problem akin to
unravelling the principles of atomic physics from spectroscopy or the
events in the early universe from cosmic radiation. In this
  paper, the statistics of gene expression dynamics is deduced from
  experimental data. Based on this statistics, a Langevin model for
  the non-equilibrium dynamics of mRNA concentrations is used to infer
  basic biophysical parameters of expression dynamics, including mRNA
  degradation constants, the statistics of transcription factor-DNA
  interactions, and regulatory interactions.

\section{Modelling gene expression dynamics}

The basic processes driving the concentration of mRNA are
transcription, which creates mRNA molecules at a certain rate, and
mRNA degradation, which destroys molecules at a rate proportional to
their abundance. This dynamics gives rise to the well-known 
linear equation of
motion~\cite{Monodetal:1952,Alonbook:2007} for the concentration $x$ of 
mRNA of a given gene 
\be
\label{deterministic1}
\partial_t x = -\eta x + f \ ,
\ee
where $\eta$ is a mRNA-specific degradation constant and $f$ is the
transcription rate. 

To model changes in the transcription rate over time, this
deterministic dynamics is generalized to a stochastic process
described
by~\cite{Paulsson:2004,Ozbudaketal:2004,ChenEnglandShakhnovich:2004}
\be
\label{langevin1}
\partial_t x \equiv \frac{x(t+\Delta_t)-x(t)}{\Delta_t}  = 
 -\eta x + f + \sqrt{D/\Delta_t} \, \xi(t) \ . 
\ee
At this point, the term $\xi(t)$ describes lack of 
knowledge regarding regulation by gene-specific transcription 
factors whose concentrations vary over time, and possible experimental
noise. We model this term by a random variable, whose mean and variance 
are zero and one, respectively, without loss of generality. Regulation will be incorporated below. 
The amplitude $\sqrt{D/\Delta_t}$ of this noise term is characterized by a 
`diffusion constant' $D$ and the interval $\Delta_t$ between 
two successive measurements. Equation~(\ref{langevin1}) is 
well-known as the Langevin equation describing an Ornstein-Uhlenbeck 
process~\cite{vanKampenbook}. 

\begin{figure}[tbh!]
\subfigure[\label{fig:delx_5014}]{
\includegraphics[width=.23\textwidth]{5014_drift.eps}}
\subfigure[\label{fig:xidist_5014}]{
\includegraphics[width=.215\textwidth]{5014_xidist.eps}}
\vspace{1ex}

\subfigure[\label{fig:delx_mean}]{
\includegraphics[width=.225\textwidth]{delq_mean_tk8.eps}}
\subfigure[\label{fig:xidist_all}]{
\includegraphics[width=.23\textwidth]{xidist_all_tk8.eps}}
\vspace{1ex}

\subfigure[\label{fig:varxi_vs_x}]{
\includegraphics[width=.25\textwidth]{varxi_vs_q.eps}}
\subfigure[\label{fig:xi_corr}]{
\includegraphics[width=.185\textwidth]{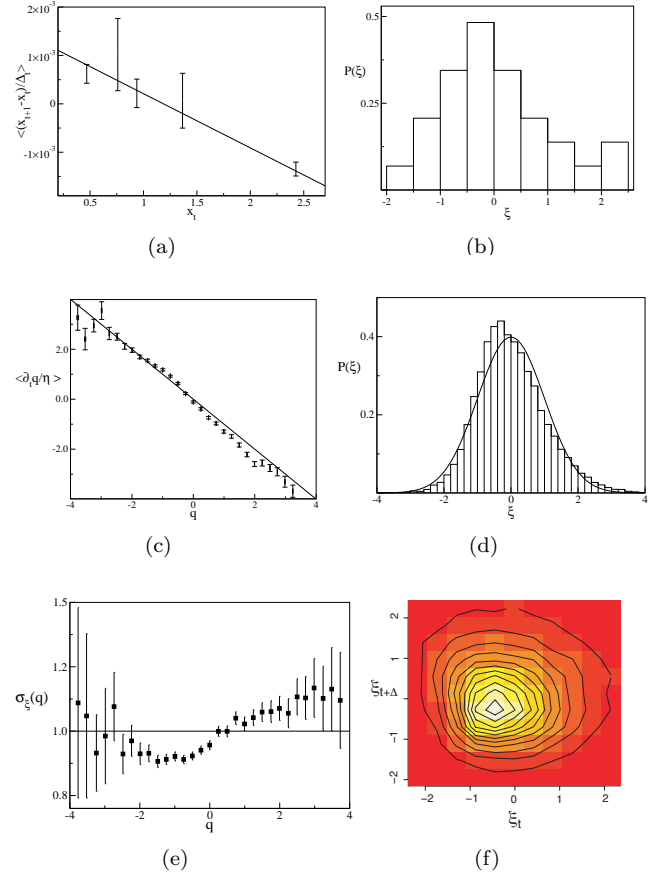}}
\caption{\label{fig:drift+diff} \small {\bf Statistics of mRNA
    concentration dynamics: Drift and diffusion.}  a) The average rate of expression level
  change, or drift $\langle(x_{t+1}-x_t)/\Delta_t \rangle$ for
  a single mRNA species is plotted against the expression levels $x_t$, by
  binning a time course of $35$ measurements into $5$ bins.  The mRNA
  chosen for this example is encoded by gene YN40, which probably
  codes for a cell wall protein~\cite{Kawahataetal:2006}.  
b) Solving
  (\ref{langevin1}) for $\xi(t)$, the statistics of the stochastic
  term can be estimated, shown here for the example gene YN40.  
c) To allow
  comparison of genes with different decay constants and transcription
  rates, rescaled expression values $q$ were used to compute the drift
  of $q$ for all genes in yeast, see text. The genome-wide result agrees well with the linear
  decay law $\langle \partial_t q /\eta \rangle = - q$ indicated by the
  solid line.  
d) The genome-wide distribution of the noise term $P(\xi)$, see text.
e) The
  standard deviation $\sigma_{\xi}(q)$ of the noise term is shown
  against the rescaled expression level $q$. 
f) The distribution
  $P(\xi_t,\xi_{t+\Delta})$ of noise terms at subsequent time
  intervals $t$ and $t+\Delta$, see text. }
\end{figure}

The soundness of (\ref{langevin1}) as a model of mRNA dynamics  
can be probed by separating deterministic
and stochastic contributions to the dynamics. The drift term
$\langle (x_{t+1}-x_t)/\Delta_t \rangle$ is computed by considering
all instances of concentration $x_t=x$ in a time course. We use the
time course data by Spellman et al.~\cite{Spellmanetal:1998}, where
expression levels of all yeast genes were measured at intervals
ranging from $7$ to $20$ minutes, see Appendix. 
The expression ratios (fluorescence measured relative to a reference
sample, also termed expression levels) were used as measures of mRNA concentration. This is valid
provided the fluorescence/concentration response is in the linear
regime~\cite{CarlonHeim:2006}.
The result for a specific mRNA shown
in Fig.~\ref{fig:delx_5014} is compatible with the linear
relationship $\langle \partial_t x \rangle=-\eta x + f$. Genome-wide
results will be discussed below. Similarly,
the distribution of the noise $\xi(t)$ can be deduced from the data, 
This distribution is shown for a single gene in 
Fig.~\ref{fig:xidist_5014} and is found
to be compatible with a Gaussian distribution. 
A Kolmogorov-Smirnov test shows that this holds also for 
all other genes.

For a Gaussian noise statistics, the equation of 
motion~(\ref{langevin1}) gives the conditional 
probability of going from concentration $x_t$ to $x_{t+1}$ in a given 
short time step as  
\bea
\label{propagator1}
\lefteqn{ G_{\eta,f,D}(x_{t+1}|x_t) }\\
&=&\frac{1}{\sqrt{2 \pi D \Delta_t}}
 \exp{\left\{ -\frac{\Delta_t}{2D} (\partial_t x + \eta x_t - f)^2 
 \right\} } \nonumber \ .
\eea
Assuming uncorrelated noise, the propagator (\ref{propagator1}) 
gives the probability for a complete time course of mRNA expression levels,
${\bf x}= (x_1,x_2,\ldots,x_T)$, starting from some 
initial point $x_1$, as 
\be
\label{likelihood1} {\mathcal P}_{\eta,f,D}({\bf x}) =
\prod_{t=1}^{T-1} G_{\eta,f,D}(x_{t+1}|x_t) \ .  
\ee 
In order to determine the model parameters we now ask
which parameter values $\eta,f,D$ give the highest likelihood
(\ref{likelihood1}) to a given time course ${\bf x}$. This leads
to the straightforward maximum likelihood estimates~\cite{Honerkampbook:1994} 
\be
\label{max_likelihood}
(\eta^{\star},f^{\star}, D^{\star}) =
{\arg\!\max}_{\eta,f,D} {\mathcal P}_{\eta,f,D}({\bf x}) \ .
\ee
The time
intervals $\Delta_t$ between adjacent time points frequently differ
from one another. This makes parameters estimates from (\ref{propagator1}) -- 
(\ref{max_likelihood}) distinct from fitting data to a deterministic
model with a quadratic penalty of 
residuals~\cite{WahdeHertz:2000,Ronenetal:2002,Khaninetal:2006}.

The experimental data~\cite{Spellmanetal:1998} gives expression level
time courses for nearly all $\sim 6000$ yeast genes. This allows to
probe the model~(\ref{langevin1}) on a genome-wide scale with high
statistical accuracy. In order to compare the dynamics of genes with
different values of the parameters $\eta,f,D$ we introduce rescaled
expression levels $q =
\sqrt{2/(D \eta)}(\eta x-f)$ so the Langevin
equation~(\ref{langevin1}) reads 
$\partial_t q= -\eta q +\sqrt{2 \eta/\Delta_t} \xi(t)$ and distribution of $q$ in equilibrium 
is independent of the parameters, $P(q) \sim \exp\{-q^2/2\}$.

The average rate of change (drift) of rescaled expression levels $q$ between
two successive time steps is shown in Fig.~\ref{fig:delx_mean} for all
genes. The linear relationship $\langle \partial_t q/\eta \rangle= -q$
is obeyed quite accurately. Similarly, the distribution of the
noise term $\xi(t)$ across all genes, shown in
Fig.~\ref{fig:xidist_all}, rather closely follows a Gaussian distribution.

At the genome-wide level a further assumption of the Langevin
equation~(\ref{langevin1}) can be examined. There the noise amplitude
was taken to be independent of the expression level $x$. To examine
this assumption, the standard deviation $\sigma_{\xi}(q)$ of the noise
term is computed at different rescaled expression levels
$q$. Fig.~\ref{fig:varxi_vs_x} shows only an increase in noise
variance of $15\%$ over the whole range of expression level
changes. This increase of the noise amplitude may be due to increased
experimental noise at higher expression levels, possibly arising from
an intensity-dependent sensitivity of the fluorescence measurement of
the microarrays. An alternative scenario for a changing
noise-amplitude are stochastic fluctuations of the degradation rate
$\eta$.  However, such fluctuations would lead to a linear
multiplicative noise term in the Langevin equation~(\ref{langevin1}),
that is, a noise amplitude which linearly increases with $q$. The
saturation of $\sigma_{\xi}(q)$ at high and low values of $q$ which is
seen in Fig.~\ref{fig:varxi_vs_x} does not support such a linear
multiplicative scenario.  This result -- a small, nonlinear change in
the noise amplitude -- contradicts stochastic models with purely
multiplicative noise which have been used in the 
literature~\cite{Chenetal:2005,ClimescuQuirk:2007,StokicHanelThurner:2007}.

A last assumption behind the Langevin equation~(\ref{langevin1}) and
the corresponding propagator~(\ref{propagator1}) is the statistical
independence of noise terms at successive time steps. The distribution
of successive noise terms $P(\xi_{t+\Delta},\xi_t)$ across all genes
is shown in Fig.~\ref{fig:xi_corr}. For small values of $\xi$,
successive noise terms are independent, leading to circular 
contours in Fig.~\ref{fig:xi_corr}. For very short intervals
$\Delta$ between measurements, not realised in currently available
data, one should expect to find correlations between the noise terms
$\xi(t)$ and $\xi(t+\Delta)$. Hence the assumption of uncorrelated
noise will not hold for arbitrarily small sampling intervals. For this 
reason, the short term propagator~(\ref{propagator1}) was used here, 
rather than the full propagator of an Ornstein-Uhlenbeck process with 
delta-correlated noise. In practice, this choice makes little difference for 
the results obtained.  

These results show that the equation of motion~(\ref{langevin1}) with
an independent Gaussian distribution of noise of constant amplitude is
a good description of the observed expression level
dynamics. Some deviations occur at high and low expression levels. The
origin of these deviations might, however, lie in the details of how
the experiments~\cite{Spellmanetal:1998} were taken. One particular
source may be non-linearities in the relationship between expression
levels and mRNA concentrations~\cite{CarlonHeim:2006}. Future
experiments, taken with different experimental technologies and at
shorter sampling times, will allow further tests and possible
modifications of the model.

\section{mRNA degradation rates}

We now discuss the biophysical parameters inferred from the expression
level time courses of different genes. The decay times $\tau =
\eta^{-1}$ inferred using (\ref{max_likelihood}) for different genes
range from $350 \mbox{s}$ to $9900 \mbox{s}$ with mean $1400 \mbox{s}$
and standard deviation $700 \mbox{s}$.  Decay times are found to
differ somewhat across genes with different function, with mRNA of
genes involved in transcription regulation being degraded faster than
average ($\overline{\tau}= 1200 \pm 100 \mbox{s}$), and those of genes
involved in metabolic processes being degraded more slowly
($\overline{\tau}= 1500 \pm 50\mbox{s}$).  A simple possible
explanation for these different decay times is the need to quickly
change concentrations of regulatory proteins in response to changing
external conditions.

These results agree with experiments~\cite{Wangetal:2002}, where mRNA
decay times were measured after halting mRNA production through a heat
shock. (This unphysiological condition may well lead to
  different decay rates than under standard conditions.) The measured
decay times range from $\tau=160\mbox{s}$ to $\tau=7800\mbox{s}$ with
mean $1700\mbox{s}$ and standard deviation $560\mbox{s}$. The
experimental study~\cite{Wangetal:2002} also notes the relatively slow
degradation of mRNA of genes involved in metabolic processes, and the
faster degradation of those involved in regulatory systems.

\section{Regulatory interactions}

The transcription of genes is controlled by the presence of
transcription factor proteins. Accordingly, the transcription rate $f$
of a target gene depends systematically on the concentration of its
transcription factors. The dynamic equations of different genes are
thus coupled by regulatory interactions. The resulting correlation of
expression levels of different genes serves as the basis for exploring
regulatory interactions.

\begin{figure}[tbh!]
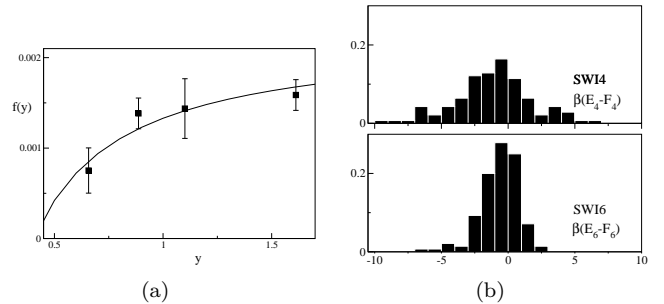

\subfigure[\label{fig:fy}]{
\includegraphics[width=.225\textwidth]{5014_fvec.eps}}
\hspace{1mm}
\subfigure[\label{fig:swi4swi6}]{
\includegraphics[width=.22\textwidth]{loghalfseqtarswi4swi6cdc28bw.eps}}
\caption{ \small {\bf Transcription regulation by transcription
    factors.} 
  a) The functional form of the transcription rate $f(y)$
  determined by discretizing the transcription factor expression levels $y$
  into four bins with equal number of incidences and inferring the
  corresponding four values of the transcription rate $f$ (solid squares) 
  along with the parameters $\eta$ and $D$. The
  result, shown here for transcription factor Swi4 and target YN40,
  agrees well with the Michaelis-Menten law~(\ref{michaelis-menten})
  displayed by the solid line. Error bars indicate the point where the
  likelihood~(\ref{likelihood1}) has decayed to $e^{-1/2}$ of its maximum,
  corresponding approximately to one standard deviation.  
  b) The distribution of binding energies (relative to their free energies in solution) of 
  the Swi4--DNA and the Swi4--Swi6 interactions, see text. 
 }
\end{figure}

We start by considering the effects of a single
transcription factor on the mRNA concentration $x$ of a given gene. The 
key questions are (i) how does the transcription rate of a target gene 
depend on the concentration of its transcription factor and (ii) can one 
identify the targets of a given transcription factor from their effect on 
the transcription rate~? We generalize the 
stochastic dynamics~(\ref{langevin1}) to 
\be 
\label{langevin2}
\partial_t x = -\eta x + f(y) + \sqrt{D/\Delta_t} \, \xi(t) \ ,
\ee
with the transcription rate $f(y)$ depending on the concentration $y$
of the transcription factor at time $t$. In the following, we will
neglect post-transcriptional regulation and take the mRNA
expression level of a transcription factor as a proxy for its protein
concentration. The functional form of $f(y)$ for a given mRNA and a
given transcription factor can be inferred by discretizing the
continuous values of $y$ into a small number of bins containing an
equal number of instances, and inferring the corresponding values of
$f$ at the centres of these bins.  An example, involving the
transcription factor Swi4 and one of its targets, gene YN40, is shown
in Fig.~\ref{fig:fy}.  Similar results are found for other
transcription factors and their targets.
%
%

Fig.~\ref{fig:fy} displays a non-linear relationship 
between transcription factor concentration and the target
transcription rate showing saturation at high concentrations. This is 
expected since at high transcription factor concentration most binding 
sites are occupied by
a transcription factor molecule, and further increases of
concentration no longer have a strong effect. Assuming that a 
transcription factor may either be bound at a binding site, or 
be suspended in solution or bound non-specifically elsewhere on DNA, 
the probability of a given site being occupied is given by elementary 
statistical mechanics~\cite{GerlandMorozHwa:2002} as  
\be
\label{binding_prob}
P(y)=\frac{y e^{-\beta E}}{y e^{-\beta E}+e^{-\beta F}} 
\ee 
and depends on the transcription factor concentration $y$, 
binding energy $E$ and a free
energy $F$ of the transcription factor in solution or bound elsewhere.

Assuming the transcription rate to depend linearly on the probability 
that the binding site is occupied at a given time one obtains 
\be
\label{michaelis-menten}
f(y)=f_0+ \delta P(y) \ , 
\ee 
where $f_0$ is a basal transcription rate in the absence of
transcription factors and $\delta$ quantifies the change of the
transcription rate due to transcription factor binding. Positive
values of $\delta$ describe enhancers and negative values of $\delta$ 
indicate repressors. The functional
form~(\ref{michaelis-menten}) is the celebrated Michaelis-Menten
kinetics, first studied in the context of enzymatic reactions
nearly a century ago~\cite{MichaelisMenten:1913} and used widely
in transcription modelling~\cite{Alonbook:2007}. Like many other targets of 
transcription factor Swi4, the relationship between transcription rate 
of target YN40 and transcription factor concentration fits 
the Michaelis-Menten model well, see Fig.~\ref{fig:fy}.

\begin{figure}[tbh!]
\vspace{1mm}
\includegraphics[width=.3\textwidth]{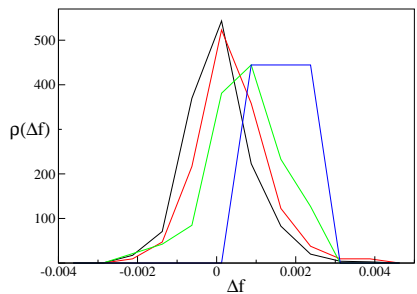}
\caption{\label{fig:deltaf}{\bf Genes with a higher number of Swi4 binding
      sites respond more strongly to changes in the Swi4
      expression level.}  The distribution of the range $\Delta f
    \equiv |f(y_{\max})-f(y_{\min})|$ of the force term is shown for
    different sets of genes: all genes (black), genes with one or more
    Swi4 binding motifs (red), genes with two or more motifs (green),
    genes with three or more Swi4 binding motifs (blue). Genes with a
    higher number of binding sites are seen to have a larger amplitude
    of the driving force.  }
\end{figure}

One can turn this result on its head and use the Michaelis-Menten
model~(\ref{michaelis-menten}) and equations~(\ref{max_likelihood}) --
(\ref{langevin2}) to infer the binding parameters
$\beta(E-F),f_0,\delta$ of a transcription factor for different
genes. We expect that genes with a high response $\Delta f
  \equiv f(y_{\max})-f(y_{\min})$ to changes in concentration of a
  transcription factor are targets of that transcription factor. 
We
  test this relationship for the transcription factor Swi4, and
  consider genes with different number of Swi4 binding sequences in
  their regulatory region, see Figure~\ref{fig:deltaf}. One finds that
  $\Delta f$ tends to increase with the number of binding sequences,
  indicating that Swi4 acts an enhancer of gene expression.
The $10$ genes with the highest $\Delta f$ are 
CDC9, RNR1, YG3N, CRH1, YIO1, RAD27, PRY2, CSI2, PMS5, and CDC21. 
With a single exception, all of these genes have at least one copy of the Swi4 binding
sequence in their regulatory region. Furthermore, $8$ of the $10$
predicted targets have been previously found experimentally~\cite{yeastract}.
  
Many genes are regulated by more than one transcription factor. Higher
organisms in particular owe much of their complexity to the
combinatorial logic implemented by the cooperative binding
of several transcription factors~\cite{BuchlerGerlandHwa:2003}. For
instance, the transcription factor Swi4 forms a dimer with
transcription factor Swi6, with Swi4 binding to regulatory DNA and
Swi6 attracting the polymerase molecule which starts
transcription~\cite{AndrewsMoore:1992}, see
Fig.~\ref{fig:transcription_schematic}. For this situation, the
binding probability (\ref{binding_prob}) generalizes to \be
  P(y_4,y_6) =
        \frac{y_4 y_6 }
        {e^{\beta (E_4-F_4+E_6-F_6)} + y_4 e^{\beta (E_6-F_6)}+y_4 y_6} \ ,
\ee
where $y_4$ and $y_6$ are the concentrations of Swi4 and Swi6, respectively. 
The constant $E_4$ gives the energy of
Swi4--DNA binding and $F_4$ the free energy of Swi4 in solution, $E_6$
is the energy of the Swi4--Swi6 interaction and $F_6$ the free
energy of Swi6 in solution. The Swi4--DNA energy depends on the
binding sequence in the regulatory region of a given gene and may
thus vary from gene to gene. On the other hand, the Swi4--Swi6
interaction does not depend on the target gene. Fig.~\ref{fig:swi4swi6}
shows the distribution of ${\beta (E_4-F_4)}$ and ${\beta (E_6-F_6)}$
inferred by maximum likelihood for different targets of Swi4--Swi6. As
expected, the variance of the Swi4--Swi6 interaction is significantly
smaller than that of the Swi4--DNA interaction.

\section{Discussion}

Some of the properties of gene expression are likely to be universal 
and occur in many organisms under many circumstances. An example is the 
biophysical implementation of 
transcription regulation through the binding of molecules, with the 
resulting saturation of bound molecules at high concentrations. Another is 
the independent degradation of mRNA molecules, leading to the linear drift
of mRNA concentrations. (This is less universal, since at very high
concentrations mRNA molecules may outnumber RNase molecules
responsible for mRNA degradation.)

Other properties of gene expression are specific to a given organism.
Indeed, higher organisms, which share largely the same set of genes, derive
most of their diversity from specific changes in how those genes
are regulated~\cite{King:1975}.  These specific properties include the
transcription factor binding sites in the regulatory region of a gene,
and the interactions between transcription factors. In this paper we
have used simple stochastic models of mRNA dynamics based on the
universal properties to infer some of the specific properties of gene
expression in yeast.  Current and future developments in molecular
biology will bring dynamically resolved data on other types of
molecules apart from messenger RNA: concentrations of regulatory
proteins themselves~\cite{Bar-Evenetal:2006,Khaninetal:2006} or of
$\mu$RNA, which silences specific mRNA molecules, leading to new
challenges in the non-equilibrium dynamics of genetic regulation.

\acknowledgements Funding from the DFG is acknowledged under 
grant BE 2478/2-1 and SFB 680. This research was supported in part by 
the National Science Foundation under Grant No. PHY05-51164. 

\section{Appendix}

The mRNA expression data of Spellman~\cite{Spellmanetal:1998}
comprises four time courses corresponding to different ways of
synchronizing yeast cell cultures to the same phase in the cell
cycle. Three timecourses were used here (termed alpha, cdc15, cdc28
in~\cite{Spellmanetal:1998}), each set equally contributing to the
likelihood~(\ref{likelihood1}). These different timecourses 
correspond to different ways of synchronizing a population of cells 
in a given phase of the cell cycle. The fourth set (elu) was not used as
it systematically lead to decreased likelihoods
in~(\ref{likelihood1}). The likelihood~(\ref{likelihood1}) also allows
to identify probes on the chip which give results compatible with
uncorrelated random data; only probes where~(\ref{likelihood1}) gave
log-likelihoods larger than $1.1$ the log-likelihood of the data under
an uncorrelated Gaussian model were used.  Set alpha contains $18$
samples spaced $7$ minutes apart, cdc15 $24$ samples $10-20$ minutes
apart, cdc28 contains $17$ samples spaced $10$ minutes apart.  

The functional classification of genes followed the Gene
Ontology~\cite{GO:2000} using the terms \textit{transcription
  regulation activity} (GO:0030528) and \textit{metabolic process}
(GO:0008152).

The canonical binding sequence for Swi4 is ``CRCGAAA'' where R stands
for either G or A~\cite{ChenHataZhang:2004}. Regions $500$
base pairs before the transcription initiation point were searched for
copies of this sequence (or its reverse, its complement, or its
reverse complement) . Sequences were taken from {\tt
  www.ncbi.nlm.nih.gov/CBBresearch/Landsman/Cell\_
cycle\_data/upstream\_seq.html}
$24\%$ of all yeast genes contain at least one copy of the Swi4
binding sequence in their regulatory region, $4\%$ contain two copies
or more.  The {\sc yeastract}-database~\cite{yeastract} list
$170$ Swi4 targets verified experimentally, or about $3\%$ of the yeast
genes.

The example gene YN40 in Figs.~\ref{fig:drift+diff} and \ref{fig:fy} 
was chosen due to the high number of $3$ Swi4 binding sites within
$500$ basepairs and $5$ sites within $1000$ basepairs of the 
transcription initiation site.


\begin{thebibliography}{10}
\expandafter\ifx\csname url\endcsname\relax\def\url#1{\texttt{#1}}\fi

\bibitem{Leeetal:2002}
\Name{Lee T.~I. {\it et al.}}
  \REVIEW{Science }{298}{2002}{799}.

\bibitem{Harbisonetal:2004}
\Name{Harbison C.~T. {\it et al.}}
  \REVIEW{Nature }{431}{2004}{99}.

\bibitem{HertzStormo:1999}
\Name{Hertz G. \and Stormo G.} \REVIEW{Bioinformatics }{15}{1999}{563}.

\bibitem{BussemakerSiggia:2000}
\Name{Bussemaker H., Li H. \and Siggia E.~D.} \REVIEW{Proc. Natl. Acad. Sci.
  USA }{97}{2000}{10096}.

\bibitem{Chuaetal:2004}
\Name{Chua G., Robinson M.~D., Morris Q. \and Hughes T.~R.} \REVIEW{Curr Opin
  Microbiol }{7}{2004}{638}.

\bibitem{GEOgene_expression}
\Book{Gene expression omnibus ({GEO})} \url{http://www.ncbi.nlm.nih.gov/geo/ }
  (2007).

\bibitem{Margolin.etal:2004}
\Name{Margolin A. {\it et al.}} \REVIEW{BMC Bioinformatics }{7}{2006}{S7}.

\bibitem{Hertz:1998}
\Name{Hertz J.} 
\url{citeseer.ist.psu.edu/hertz98statistical.html} (1998).

\bibitem{Friedman:2004}
\Name{Friedman N.} \REVIEW{Science }{303}{2004}{799}.

\bibitem{BarJoseph:2004}
\Name{{Bar-Joseph}, Z.} \REVIEW{Bioinformatics }{20}{2004}{2493}.

\bibitem{Lebre:2007}
\Name{L{\`e}bre S.} 
\url{http://arxiv.org/abs/0704.2551v1} (2007).

\bibitem{ChuGlymourScheinesSpirtes:2003}
\Name{Chu T., Glymour C., Scheines R. \and Spirtes P.} \REVIEW{Bioinformatics
  }{19}{2003}{1147}.

\bibitem{Monodetal:1952}
\Name{Monod J., {Pappenheimer, Jr} A. \and Cohen-Bazire G.} \REVIEW{Biochim.
  Biophys. Acta }{9}{1952}{648}.

\bibitem{Alonbook:2007}
\Name{Alon U.} \Book{An Introduction to System Biology: Design Principles of
  Biological Circuits} (Chapman \& Hall, Boca Raton, FL) 2007.

\bibitem{Paulsson:2004}
\Name{Paulsson J.} \REVIEW{Nature }{427}{2004}{415}.

\bibitem{Ozbudaketal:2004}
\Name{Ozbudak E., Thattai M., Kurtser I., Grossman A. \and {van Oudenaarden}
  A.} \REVIEW{Nature Genetics }{31}{2002}{69}.

\bibitem{ChenEnglandShakhnovich:2004}
\Name{Chen W., England J. \and Shakhnovich E.} 
\url{http://arxiv.org/abs/q-bio.MN/0402021} (2004).

\bibitem{vanKampenbook}
\Name{van Kampen N.} \Book{Stochastic Processes in Physics and Chemistry}
  (Elsevier Science, Amsterdam) 1992.

\bibitem{Spellmanetal:1998}
\Name{Spellman P.~T., {\it et al.}} \REVIEW{Mol. Biol. Cell
  }{9}{1998}{3273}.

\bibitem{CarlonHeim:2006}
\Name{Carlon E. \and Heim T.} \REVIEW{Physica A}{362}{2006}{433}.

\bibitem{Kawahataetal:2006}
\Name{Kawahata M., Masaki K., Fujii T. \and Iefuji H.} \REVIEW{FEMS Yeast Res
  }{6}{2006}{924}.

\bibitem{Honerkampbook:1994}
\Name{Honerkamp J.} \Book{Stochastic Dynamical Systems: Concepts, Numerical
  Methods, Data Analysis} (VCH, New York) 1994.

\bibitem{WahdeHertz:2000}
\Name{Wahde M. \and Hertz J.} \REVIEW{Biosystems }{55}{2000}{126}.

\bibitem{Ronenetal:2002}
\Name{Ronen M., Rosenberg R., Shraiman B.~I. \and Alon U.} \REVIEW{Proc Natl
  Acad Sci U S A }{99}{2002}{10555}.

\bibitem{Khaninetal:2006}
\Name{Khanin R., Vinciotti V. \and Wit E.} \REVIEW{Proc. Natl. Acad. Sci. USA
  }{103}{2006}{18592}.

\bibitem{Chenetal:2005}
\Name{Chen K., Wang T., Tseng H., Huang C. \and Kao C.} \REVIEW{Bioinformatics
  }{21}{2005}{2883}.

\bibitem{ClimescuQuirk:2007}
\Name{Climescu-Haulica A. \and Quirk M.~D.} \REVIEW{BMC Bioinformatics
  }{8}{2007}{S4}.

\bibitem{StokicHanelThurner:2007}
\Name{Stokic D., Hanel R. \and Thurner S.} 
\url{http://arxiv.org/abs/0711.4775}, (2007).

\bibitem{Wangetal:2002}
\Name{Wang Y., {\it et al.}} \REVIEW{Proc. Natl. Acad. Sci. USA }{99}{2002}{5860}.

\bibitem{GerlandMorozHwa:2002}
\Name{Gerland U., Moroz D. \and Hwa T.} \REVIEW{Proc.Nat. Acad. Sci. USA
  }{99}{2002}{12015}.

\bibitem{MichaelisMenten:1913}
\Name{Michaelis L. \and Menten M.} \REVIEW{Biochem. Z. }{49}{1913}{333}.

\bibitem{yeastract}
\Book{{YEASTRACT}} \url{http://www.yeastract.com/} (2007).

\bibitem{BuchlerGerlandHwa:2003}
\Name{Buchler N., Gerland U. \and Hwa T.} \REVIEW{Proc. Natl. Acad. Sci. USA
  }{100}{2003}{5136}.

\bibitem{AndrewsMoore:1992}
\Name{Andrews B.~J. \and Moore L.~A.} \REVIEW{Proc Natl Acad Sci U S A
  }{89}{1992}{11852}.

\bibitem{King:1975}
\Name{King M. \and Wilson A.} \REVIEW{Science }{188}{1975}{107}.

\bibitem{Bar-Evenetal:2006}
\Name{Bar-Even A.,{\it et al.}} \REVIEW{Nature Genetics }{38}{2006}{636}.

\bibitem{GO:2000}
\Name{{The Gene Ontology Consortium}} \REVIEW{Nature Genet. }{25}{2000}{25}.

\bibitem{ChenHataZhang:2004}
\Name{Chen G., Hata N. \and Zhang M.} \REVIEW{Nucleic Acids Res.
  }{32}{2004}{2362}.

\end{thebibliography}
\end{document}